\documentclass[twocolumn,prb,superscriptaddress,showpacs,epsf]{revtex4}
\usepackage[latin9]{inputenc}
\usepackage{amsmath}
\usepackage{amssymb}
\usepackage{graphicx}

\makeatletter
\@ifundefined{textcolor}{}
{%
 \definecolor{BLACK}{gray}{0}
 \definecolor{WHITE}{gray}{1}
 \definecolor{RED}{rgb}{1,0,0}
 \definecolor{GREEN}{rgb}{0,1,0}
 \definecolor{BLUE}{rgb}{0,0,1}
 \definecolor{CYAN}{cmyk}{1,0,0,0}
 \definecolor{MAGENTA}{cmyk}{0,1,0,0}
 \definecolor{YELLOW}{cmyk}{0,0,1,0}
}

\usepackage{epstopdf}

\usepackage{rotating}
\usepackage{soul}\usepackage{color}

\makeatother

\begin{document}

\title{Superconducting properties of corner-shaped Al microstrips}

\author{O.-A. Adami\footnote{These authors contributed equally to this work.}}

\affiliation{Département de Physique, Université de Liège, B-4000 Sart Tilman,
Belgium}

\author{D. Cerbu$^{*}$}

\affiliation{INPAC -- Institute for Nanoscale Physics and Chemistry, Nanoscale
Superconductivity\\
 and Magnetism Group, K.U.Leuven, Celestijnenlaan 200D, B--3001 Leuven,
Belgium}

\author{D. Cabosart}

\affiliation{NAPS/IMCN, Université catholique de Louvain, B-1348 Louvain-la-Neuve,
Belgium}

\author{M. Motta}

\affiliation{Departamento de Física, Universidade Federal de São Carlos, 13565-905
São Carlos, SP, Brazil}

\author{J. Cuppens}

\affiliation{INPAC -- Institute for Nanoscale Physics and Chemistry, Nanoscale
Superconductivity\\
 and Magnetism Group, K.U.Leuven, Celestijnenlaan 200D, B--3001 Leuven,
Belgium}

\author{W.A. Ortiz}

\affiliation{Departamento de Física, Universidade Federal de São Carlos, 13565-905
São Carlos, SP, Brazil}

\author{V. V. Moshchalkov}

\affiliation{INPAC -- Institute for Nanoscale Physics and Chemistry, Nanoscale
Superconductivity\\
 and Magnetism Group, K.U.Leuven, Celestijnenlaan 200D, B--3001 Leuven,
Belgium}

\author{B. Hackens}

\affiliation{NAPS/IMCN, Université catholique de Louvain, B-1348 Louvain-la-Neuve,
Belgium}

\author{R. Delamare}

\affiliation{ICTEAM, Université catholique de Louvain, 1348 Louvain-la-Neuve,
Belgium}

\author{J. Van de Vondel}

\affiliation{INPAC -- Institute for Nanoscale Physics and Chemistry, Nanoscale
Superconductivity\\
 and Magnetism Group, K.U.Leuven, Celestijnenlaan 200D, B--3001 Leuven,
Belgium}

\author{A. V. Silhanek}

\affiliation{Département de Physique, Université de Liège, B-4000 Sart Tilman,
Belgium}

\date{\today}
\begin{abstract}
The electrical transport properties of corner-shaped Al superconducting
microstrips have been investigated. We demonstrate that the sharp
turns lead to asymmetric vortex dynamics, allowing for easier penetration
from the inner concave angle than from the outer convex angle. This
effect is evidenced by a strong rectification of the voltage signal
otherwise absent in straight superconducting strips. At low magnetic
fields, an enhancement of the critical current with increasing magnetic
field is observed for a particular combination of field and current
polarity, confirming a recently theoretically predicted competing
interplay of superconducting screening currents and applied currents
at the inner side of the turn.
\end{abstract}
\maketitle

The ability of superconductors to carry electricity without resistance
holds in a restricted current density range $j<j_{max}$. Several
physical mechanisms can be identified as responsible for limiting
$j_{max}$ such as the motion of vortices, the formation of phase
slip centers or eventually when the pair-breaking current, $j_{pb}$,
is reached.

In principle, it is possible to attain the ultimate limit $j_{max}$=$j_{pb}$
by properly choosing the dimensions of the superconducting strip.
Indeed, if the width $w$ of the strip is such that $w<4.4\xi$, where
$\xi$ is the coherence length, vortices cannot fit into the sample\cite{likharev}
and therefore $j_{max}$ cannot be limited by a vortex depinning process.
In 1980 Kupriyanov and Lukichev \cite{kupriyanov} were able to determine
theoretically $j_{pb}$ for all temperatures, by solving the Eilenberger
equations, and only two years later their predictions were experimentally
confirmed by Romijn \textit{et al.}\cite{romijn} using \textit{straight}
Al strips. These works focused on the case where $w\ll\Lambda$, with
$\Lambda=2\lambda^{2}/d$ the Pearl length\cite{pearl}, $\lambda$
the London penetration depth, and $d$ the thickness of the superconductor.

Recently a renewed interest for understanding the limiting factors
of $j_{max}$ in \textit{non-straight} strips has arisen, partially
motivated by the ubiquitous presence of sharp turns in more realistic
architectures as those used in the superconducting meanders for single
photon and single electron detectors \cite{bulaevski,victor}.

Early theoretical calculations by Hagerdorn and Hall\cite{hall} showed
that a sharp bend in a superconducting wire leads to current crowding
effects at the inner corner of the the bend, which in turn reduces
the total critical current when compared to a straight wire. Not only
sharp angles along the superconducting bridge, but any sudden change
in the cross section of the wire, can lead to a reduction of the critical
current. For instance, it has been pointed out in Ref.{[}\onlinecite{silhanek-comment}{]}
that a sudden increase in the cross section of a transport bridge
leads to severe modifications of the voltage-current characteristics
rendering unreliable those measurements performed in cross-shaped
geometries. More recently, Clem and Berggren\cite{clem-berggren}
have theoretically demonstrated that sudden increases in the cross
section of a transport bridge, as those caused by voltage leads, also
produce current crowding effects and the consequent detriment of the
critical current, similarly to right-angle bends. These predictions
have been independently confirmed experimentally by Hortensius \textit{et
al.}\cite{hortensius} and by Henrich \textit{et al.}\cite{henrich}
in submicron scale samples of NbTiN and NbN, respectively, and found
to be also relevant in larger samples\cite{vestgarden}.

The effect of a magnetic field applied perpendicularly to the plane
containing the superconducting wire with a sharp turn has been discussed
in Ref.{[}\onlinecite{hortensius}{]} and Ref.{[}\onlinecite{clem-peeters}{]}.
Strikingly, in Ref.{[}\onlinecite{clem-peeters}{]} it is theoretically
predicted that due to compensation effects between the field induced
stream-lines and the externally applied current at the current crowding
point, the critical current of thin and narrow superconducting bridges
($\xi\ll w\ll\Lambda$) should \textit{increase} with field for small
fields values and for a particular polarity of the applied field.

In this work we provide experimental confirmation of the theoretical
predictions of Ref.{[}\onlinecite{clem-peeters}{]} and show that
current crowding leads also to a clearly distinct superconducting
response for positive and negative fields (or currents), making these
asymmetric superconducting nanocircuits potentially efficient voltage
rectifiers.


The samples investigated were all co-fabricated on the same chip and
consist of electron-beam lithographically defined Al structures of
thickness $d$ = 67 $\pm$ 2 nm, deposited by rf sputtering on top
of a Si/SiO$_{2}$ substrate. We focus on two different geometries.
Sample S90 consist of a 3.3 $\mu$m wide transport bridge with a 90$^{\circ}$
corner equidistant from two voltage probes separated 9.6 $\mu$m from
the inner angle of the sharp bend. Similarly, S180 is a conventional
straight transport bridge 3.7 $\mu$m wide and with voltage probes
separated by 20.9 $\mu$m. These dimensions depart from the nominal
values and were obtained via atomic force microscopy as shown in Figure
1(a)-(b).

The field dependence of the superconducting-to-normal metal transitions,
$T_{c}$($H$), determined as 0.95$R_{N}$, where $R_{N}$ is the
normal state resistance, and using an ac-current\cite{electronics}
of 1 $\mu$A, is basically the same for the two samples studied (see
Figure 1(c)). This similarity of the phase boundaries allows us to
make reliable and direct comparisons between the two samples without
the necessity to work with reduced temperatures or field units. The
critical temperature at zero field is $T_{c0}=1.320\pm0.008$ K and
the superconducting coherence length obtained from the Ginzburg-Landau
approximation is $\xi(0)=121\pm3$ nm. The BCS coherence length for
Al of similar characteristics \cite{romijn} ($T_{c0}$ and $d$)
as the one used here is $\xi_{0}=1320$ nm, indicating that our Al
falls in the dirty limit $\ell\ll\xi_{0}$, with $\ell$ the electronic
mean free path. Using the relation $\xi(0)=0.855\sqrt{(}\xi_{0}\ell)$
we deduce $\ell\sim15$ nm. An independent estimation of $\ell\sim17$
nm can be obtained from the normal state resistivity $\rho=2.0\pm0.1^{.}10^{-8}$
$\Omega$m, and taking\cite{romijn} $\rho\ell=4^{.}10^{-16}$ $\Omega$m$^{2}$.
In the dirty limit the magnetic penetration depth is given by $\lambda(0)=\lambda_{L}(0)\sqrt{\xi_{0}/\ell}\approx$
145 nm, where $\lambda_{L}(0)=$16 nm is the London penetration depth.
For thin film geometry with a perpendicular external field we need
to use the Pearl length\cite{pearl} $\Lambda=2\lambda^{2}/d$. In
the considered samples $\Lambda>2w$ for $T>$1.19 K.

\begin{figure}
\centering \includegraphics[width=7cm]{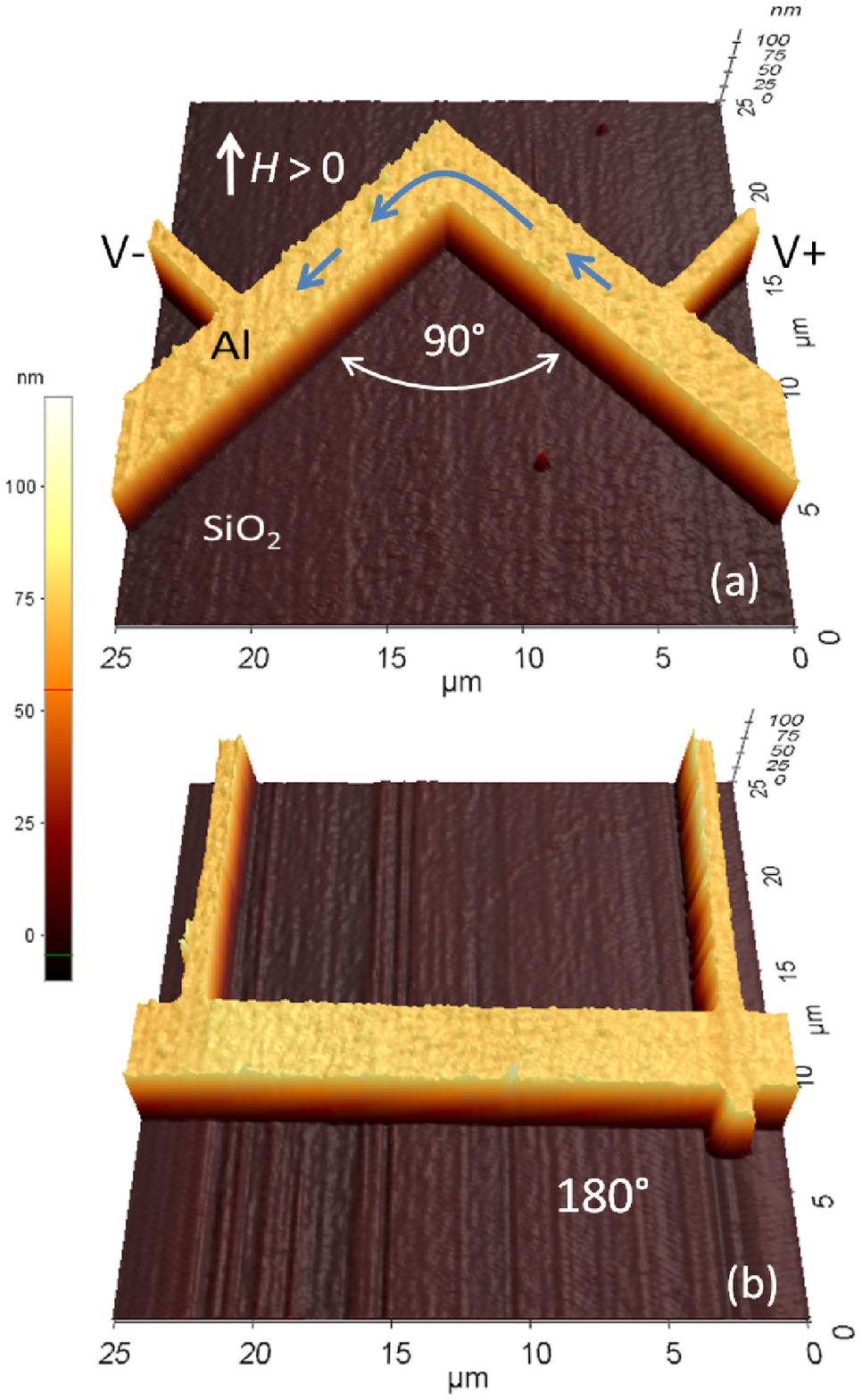} \includegraphics[width=7cm]{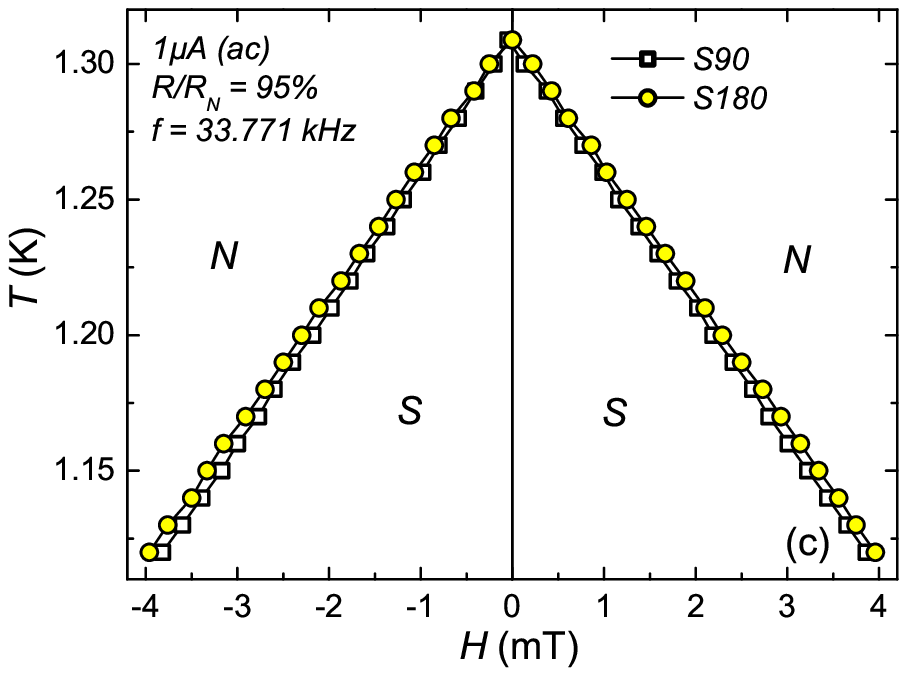}
\caption{(Color online) Atomic force microscopy images of the two superconducting
Al bridges studied: (a) S90, and (b) S180. Panel (c) shows the superconducting(S)-normal(N)
$H-T$ phase diagrams for both samples}
\end{figure}

Let us now concentrate on the current-voltage characteristics, $V$($I$),
of the considered systems. At zero external field, the $V$($I$)
curves and, in particular the critical current, $I_{c}$, should be
uniquely defined, irrespective of the direction of the applied current.
This independence on the direction of the current persists at all
fields for the S180 sample, but does not hold for the S90 sample.
Indeed, on the one hand, the outer angle of the sharp corner has a
larger surface nucleation critical field $H_{c3}$ (a factor $\sim1.16$
higher for the S90) when compared to the critical field at the inner
corner\cite{schweigert} thus making the outer corner a point of enhanced
superconductivity\cite{victor}. On the other hand, stream-lines of
the applied current tend to conglomerate at the inner corner\cite{hall},
depleting the order parameter at that place. Notice that both effects,
larger surface nucleation field and lower applied current density
at the sharper corner, share the same origin in the impossibility
of both, screening or applied currents, to reach the tip of the bend.

The fact that current crowding at the inner corner leads to local
depletion of the superconducting order parameter implies automatically
a reduction of the surface barrier for vortex penetration\cite{clem-peeters}
as long as the applied current is such that the Lorentz force pushes
vortices from the inner towards the outer corner. However, if the
current is reversed, vortices will not penetrate from the outer corner
(where total current is nearly zero) but rather symmetrically from
the straight legs of the bridge\cite{clem-peeters}. As a consequence
of this different nucleation position and nucleation condition for
the two opposite current directions, it is predicted that such a simple
corner shape wire will give rise to asymmetric $V$($I$) characteristics
and therefore to a vortex ratchet effect.

In order to demonstrate the existence of vortex motion rectification
we submitted the samples to an ac current excitation of zero mean,
$I_{ac}$, while measuring simultaneously the dc drop of voltage $V_{dc}$.
The results of these measurements $V_{dc}$($I_{ac}$) are presented
in Figure 2 for both samples. The chosen temperature $T=1.22$ K is
such that $4.4\xi=1.9\mu$m $<$ $w=3\mu$m $<$ $\Lambda=8.3\mu$m
ensuring the existence of vortices within the superconductor. There
are several points that deserve to be highlighted here, (i) rectification
effects are almost completely absent in the S180 sample, (ii) there
is a very strong ratchet signal for the S90 sample, (iii) the ratchet
signal changes polarity at zero field. Ideally, we expect no ratchet
effect at all from the S180 sample, however, the fact that both voltage
contacts are on the same side of the strip already impose a weak asymmetry
in the system which can lead to asymmetric vortex penetration\cite{vodolazov-peeters,cerbu}.
In any case, the rectification signal obtained in the S180 sample
is negligible in comparison to that observed in the sample with the
sharp turn. The fact that the rectification signal is positive at
positive fields for the S90 sample, and according to the sign convention
depicted in Fig. 1(a), we conclude that the easy direction of vortex
flow is from the inner corner towards the outer corner, in agreement
with the theoretical findings \cite{clem-peeters}. In Fig.3 we show
how the ratchet signal progressively disappears as the temperature
approaches 1.280 K. For temperatures above this value vortices cannot
fit anymore in the bridge and consequently the difference between
the two corners vanishes. Similar ratchet effects due to surface barrier
asymmetry, have been recently reported\cite{kajino} in high-Tc superconducting
asymmetric nanobridges, with one side straight and the other having
a constriction with an angle of 90$^{\circ}$.

\begin{figure}
\centering \includegraphics[width=8cm]{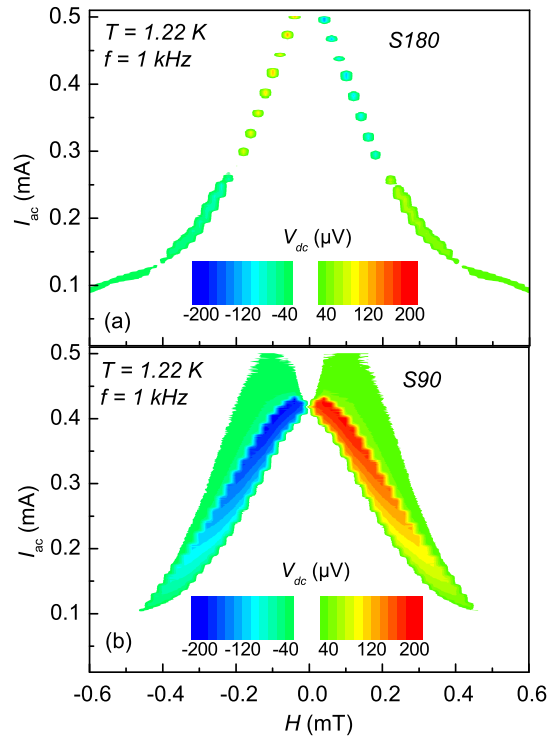} \caption{(Color online) Contour plot of the dc voltage $V\-(dc)$ as a function
of magnetic field and ac current amplitude at $T$ = 1.220 K, and
frequency of 1 kHz for sample S180 (a) and sample S90 (b).}
\end{figure}

\begin{figure}
\centering \includegraphics[width=8cm]{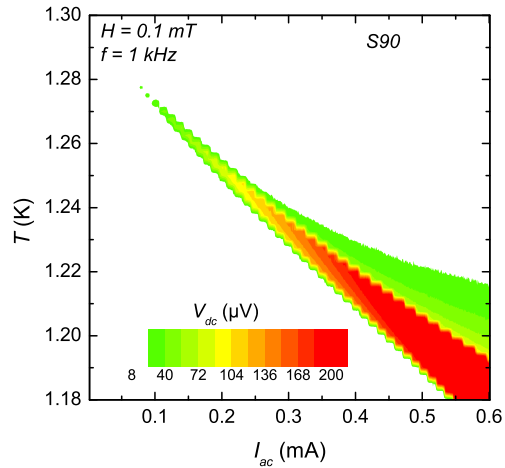} \caption{(Color online) Contour plot of the dc voltage $V\-(dc)$ as a function
of temperature and ac current amplitude at $H$ = 0.1 mT, and frequency
of 1 kHz for the sample S90.}
\end{figure}

Notice that the ratchet effect here described results from the crowding
of the applied current at the inner corner, and it would exist even
if no screening currents were present. Let us now consider the additional
effect of the screening currents. As it has been pointed out in Ref.{[}\onlinecite{clem-peeters}{]}
based on both, London and Ginzburg-Landau theories, for a given direction
of the applied current (as indicated in Fig.1(a)) a positive magnetic
field will reinforce the total current (i.e. applied plus screening)
at the inner corner and therefore the critical current will decrease
as the field intensity increases. On the contrary, a negative applied
magnetic field will induce a screening current which partially compensates
the applied current at the inner corner and a field dependent increase
of the critical current is expected \cite{clem-peeters}. We have
experimentally confirmed this prediction by measuring the critical
current using a voltage criterion of 1 $\mu$V as a function of field
and current orientation. The results are presented in Figure 4(a)
for three different temperatures and for the case where $\xi<w<\Lambda$.
For positive current and field (as defined in Fig.1(a)), we observe
a monotonous decrease of $I_{c}$. In contrast to that, for positive
current and negative field, a clear enhancement of $I_{c}$ with field
is observed for $H<H_{max}$, whereas for $H>H_{max}$ a monotonous
decrease of $I_{c}$ is recovered as a consequence of antivortices
induced by the magnetic field\cite{clem-peeters} that start to penetrate
the sample. Reversing the applied current should lead to the opposite
behavior, as indeed observed in Fig.4(a). This double test for all
polarities of current and field also permits us to accurately determine
the value of zero external field at the point where both curves cross
each other. This has been convincingly confirmed by independent measurement
of the remanent field in the S180 sample. For the sake of comparison,
in the inset of Fig.4(b) we show the critical current for the S180
sample as a function of field. Notice that for this sample, the peak
of maximum critical current is located at $H=0$, in contrast to the
behavior observed in sample S90. It is important to point out that in Ref.[13] the theoretical prediction of the curves in Fig.4(a) corresponds to a sharp inverted-V shape according to the London model, whereas the Ginzburg-Landau calculations yield a rounded top, which becomes sharper the smaller the ratio of $\xi$ to $w$. This effect appears to be confirmed, at least qualitatively, in Fig. 4(a), in which the peaks become more rounded as the temperature increases and $\xi$ increases. 

The compensation field $H_{max}$ is expected to depend on temperature
since it is determined by the screening currents. In Fig.4(b) we plot
the temperature dependence of $H_{max}/H_{c2}(T)$ where it can be
noticed that this compensation field $H_{max}$ is a small fraction
of the upper critical field $H_{c2}(T)$ in agreement with the theoretical
calculations\cite{clem-peeters}.

\begin{figure}
\centering \includegraphics[width=9.5cm]{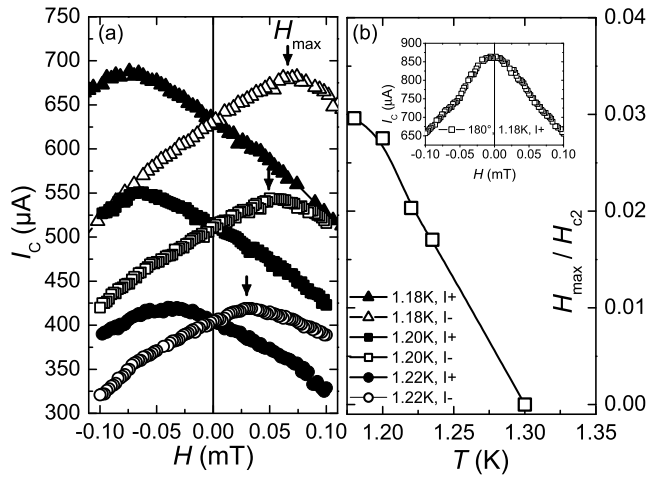} \caption{The critical current $I_{c}$ of sample S90 as a function of applied
magnetic field for both polarities of the applied current (a). Panel
(b) shows the maximum field $H_{max}$(normalized to the critical
magnetic field $H_{c2}$) as a function of temperature. The inset
in (b) shows for comparison the critical current of sample S180 versus
magnetic field at positive applied currents.}
\end{figure}


To summarize, the superconducting properties of corner-shaped Al microstrips
have been investigated. We show that sharp 90 degrees turns lead to
asymmetric vortex penetration, being easier for vortices to penetrate
from the inner side than from the outer side of the angle. We provide
experimental confirmation of the predicted\cite{clem-peeters} competing
interplay of superconducting screening currents and applied currents
at the inner side of the turn. We prove that current crowding leads
to a distinctly different superconducting responses for positive and
negative fields (or currents). These effects are evidenced also by
a field dependent critical current enhancement and also by a strong
rectification of the voltage signal, thus making these asymmetric
superconducting nanocircuits efficient voltage rectifiers. Complementary
measurements done in samples with 30$^{\circ}$ and 60$^{\circ}$
corners (not shown) reproduce the results presented here, i.e. ratchet
signal and field-induced increase of critical current.

This work was partially supported by the Fonds de la Recherche Scientifique
- FNRS, FRFC grant no. 2.4503.12, the Methusalem Funding of the Flemish
Government, the Fund for Scientific Research-Flanders (FWO-Vlaanderen),
the Brazilian funding agencies FAPESP and CNPq, and the program for
scientific cooperation F.R.S.-FNRS-CNPq. J. V.d.V. acknowledges support
from FWO-Vl. The authors acknowledge useful discussions with V. Gladilin.


\begin{thebibliography}{10}
\bibitem{electronics} The transport measurements have been done with
the sample immersed in superfluid $^{4}$He for minimizing heating
effects. Special care has been taken to avoid the high frequency noise
signal (above $\sim$ 1 MHz ) by using a pi-filter.
\end{thebibliography}
\end{document}